\documentclass[twocolumn]{aastex61}

\usepackage{graphicx}
\usepackage{amsmath}
\usepackage{mathptmx}
\usepackage{times}
\usepackage{epsfig}
\usepackage{color}
\usepackage{mathrsfs}

\begin{document}

\title{Formation of black hole X-ray binaries with non-degenerate donors in globular clusters}

\author[0000-0001-6251-5315]{Natalia Ivanova}
\affiliation{Department of Physics, University of Alberta, Edmonton, AB, T6G 2E7, Canada}
\affiliation{Kavli Institute for Theoretical Physics, UCSB, Santa Barbara, CA 93106, USA}

\author{Cassio A.~da~Rocha}
\affiliation{Department of Physics, University of Alberta, Edmonton, AB, T6G 2E7, Canada}
\affiliation{Instituto de F\'isica, Universidade Federal de Goi\'s, Goi\^ania, Goi\'as, 74.690-900, Brazil}

\author{Kenny X.~Van}
\affiliation{Department of Physics, University of Alberta, Edmonton, AB, T6G 2E7, Canada}
\affiliation{Kavli Institute for Theoretical Physics, UCSB, Santa Barbara, CA 93106, USA}

\author{Jose L.A.~Nandez}
\affiliation{Department of Physics, University of Alberta, Edmonton, AB, T6G 2E7, Canada}
\affiliation{SHARCNET, Faculty of Science, University of Western Ontario, London, ON, N6A 5B7, Canada}

\correspondingauthor{Natalia Ivanova}
\email{nata.ivanova@ualberta.ca}

\begin{abstract}
  In  this {\it  Letter} we propose a formation channel for low-mass X-ray binaries with black hole accretors and non-degenerate donors via grazing tidal encounters with subgiants. We estimate that in a typically dense globular cluster with the core density of $10^5$ stars pc$^{-3}$, the formation rates are about one binary per Gyr per 50-100 retained black holes.  The donors  -- stripped subgiants -- will be strongly underluminous when compared to subgiant or giant branch stars of the same colors.  The products of tidal stripping are underluminous by at least one magnitude for several hundred million years, when compared to normal stars of the same color, and differ from underluminous red stars that could be produced by non-catastrophic mass transfer in an ordinary binary. The dynamically formed binaries become quiescent low-mass X-ray binaries, with lifetimes of about a Gyr. The expected number of X-ray binaries is one per 50-200 retained black holes, while the expected number of strongly underluminous subsubgiant is about half this. The presence of strongly underluminous stars in a globular cluster may be indicative of black holes presence. 
\end{abstract} 


\section{Introduction}
\label{sec:intro}

Radio observations of Milky Way globular clusters (GCs) have indicated
that some contain faint potential black-hole (BH) low-mass X-ray binaries  (LMXBs).
Several BH candidate binaries are under investigation -- two LMXBs in M22
\citep{2012Natur.490...71S}, one in M62 \citep{2013ApJ...777...69C},
one in 47 Tuc \citep{2015MNRAS.453.3918M}.

In two cases, white-dwarf (WD) companions are either determined to be plausible (in M22), or are
confirmed \citep[in 47~Tuc,][]{2017MNRAS.467.2199B}. 
This is consistent with an extrapolation that can be made using the 
observations of very bright extragalactic GC~LMXBs, where 
$L_x\ga10^{39}$ erg s$^{-1}$ indicates possible BH accretors.
The observational frequency of such very bright extragalactic GC~LMXBs is
about $(0.7-2)\times10^{-9}$ per $M_\odot$ in massive GCs with an average $M_{\rm V}\approx-9$
\citep{2006ApJ...647..276K,2007ApJ...660.1246S,2008ApJ...689..983H,2013ApJ...764...98K}.
An ultra-compact X-ray BH-WD binary, once formed,
will remain as a faint BH-WD LMXB for  $\sim10^4$
times longer than a very bright BH-WD LMXB \citep{2010ApJ...717..948I,2016IAUS..312..203I}.  
This implies that in a dense and massive Milky Way ~GCs,
one faint BH-WD X-ray binary can be expected per $\sim10^{5}M_\odot$.

However, for faint BH-LMXBs with {\it non-degenerate} companions, there is
no clue regarding their  formation frequency from extragalactic observations.
There is also a lack of theoretical formation scenarios.
A tidal capture of a main sequence star by a BH was previously ruled out,
as this should produce a bright source which has not yet been detected  \citep{2004ApJ...601L.171K}.
An exchange encounter between a binary and a BH would make binaries
too wide to start mass transfer (MT), if MT starts 
the duty cycle would be too low to ever be detected \citep{2004ApJ...601L.171K}.

In this {\it Letter}, we propose to examine the formation of a BH-LMXB with an underluminous red donor,
which is produced by a grazing tidal encounter of a BH and a low-mass star that recently 
evolved off its main sequence. The formation of the binary
proposed in this manuscript  is different from
the formation via physical collisions considered in \cite{2010ApJ...717..948I}
and elsewhere before, where
the entire envelope of the subgiant star is lost, and, in the
formed binary, the BH has a degenerate companion.

\section{Dynamical formation}

\subsection{The encounter rate}

\label{sec:est}

$\sigma_{\rm RG,BH}$ is the  cross-section of an enhanced by gravitational focusing
 encounter between a red giant (RG) and a BH (here we include subgiants):

\begin{eqnarray}
\sigma_{\rm RG,BH}&=&\pi r_{\rm enc}^2\left(1+\frac{2G(M_{\rm RG}+M_{\rm BH})}{r_{\rm enc}v_{\rm \infty}^2}\right)\\&\approx &\pi r_{\rm enc}^2\left(1+3800 \frac{(M_{\rm RG}+M_{\rm BH})/M_\odot}
{r_{\rm enc}/R_\odot(v_{\rm \infty}/10{\rm km/s})^2} \right)
\label{eq:sigma}
\end{eqnarray}
\noindent Here $r_{\rm enc}$ is the largest closest approach that would lead to a
specific strong encounter, and $v_\infty$ is the relative velocity at infinity.

We search for an encounter such that at periastron the stars pass relatively close, and it is possible to damp enough of the  initial kinetic energy into the non-degenerate star to form a bound binary,
$r_{\rm enc}\le r_{\rm form}$.
On the other hand, an encounter should be not too close, so stars neither collide, nor the envelope
is removed completely during the encounter, where only a naked core that would become later a WD is left, $r_{\rm enc} \ge r_{\rm wd}$. 

For the range of closest approaches $r_{\rm wd}<r_{\rm enc}\le r_{\rm form}$
(we will determine the values in \S\ref{sec:3d}),  and assuming that  $r_{\rm form}$ is of order of 
several solar radii, while $v_{\rm \infty}$ in GCs is of order of 10 km/s, we obtain:

\begin{equation}
\sigma_{\rm RG,BH}^{\rm form}\approx 1.2\times10^{4}{R_\odot}^2\frac{M_{\rm RG}+M_{\rm BH}}{M_\odot}\left(\frac{10{\rm km/s}}{v_{\rm \infty}}\right)^2\frac{ r_{\rm form}-r_{\rm wd}}{R_\odot}.
\end{equation}

\noindent The total number of encounters leading to this kind of binary formation, per BH, and per unit of time:

\begin{equation}
N^{\rm form}=n_{\rm RG}\sigma_{\rm RG,BH}^{\rm form}v_{\infty},
\end{equation} 
where $n_{\rm RG}=f_{\rm RG}n_{\rm c}$ is the number density of RGs, $f_{\rm RG}$ is the fraction of RGs in the stellar population on a GC core, typically 8-10 percent \citep{2005ApJ...621L.109I}, and  $n_{\rm c}$ is the number density in the core of a GC. Scaling to a typical massive GC where core's number density is $10^5$ per $pc^3$, and using $n_5=n_{\rm c}/10^5$, we have:

\begin{equation}
N^{\rm form}=6\times10^{-3}f_{\rm RG}n_{\rm 5} 
\frac{M_{\rm RG}+M_{\rm BH}}{M_\odot}\frac{10{\rm km/s}}{v_{\rm \infty}}
\frac{r_{\rm form}-r_{\rm wd}}{R_\odot}{\rm Gyr}^{-1},
\end{equation}

A preliminary estimate of at which distance this capture could take place can be 
done using fitting functions for the tidal energy dissipated during a stellar encounter, $E_{\rm tide}(r_{\rm p})$ \citep{1993A&A...280..174P}. Here $r_{\rm p}$ is the closest initial approach.
To form a bound binary, the initial relative kinetic energy of the two stars $E_{\rm kin}$ has to be damped. 
This defines the maximum closest approach $r_{\rm form}$ at which the encounter leads to a binary formation, $E_{\rm tide}(r_{\rm form})>E_{\rm kin}$.
As we are interested only in cases when a part of the envelope is left, the amount
of energy that must dissipate should not exceed (by absolute value) RG envelope's
binding energy $E_{\rm env}$.
This defines the minimum closest approach $r_{\rm wd}$ at which the encounter
does not fully eject the envelope, 
$E_{\rm tide}(r_{\rm wd})+E_{\rm env} < 0$.  

For a preliminary estimate, we use analytic formulae for stellar evolution 
\citep[][]{2000MNRAS.315..543H}, and  adopt that 
$E_{\rm env}=-GM_{\rm RG,env}M_{\rm RG}/r_{\rm RG}$, where $M_{\rm RG,env}$ 
is the RG envelope mass,  and $r_{\rm RG}$ is the radius of the RG.
At 10 Gyr, the RG population is formed by stars of nearly identical initial mass. The probability of the presence (at this moment of time) of giants that have radii within a specific range is hence about the same as the ratio of how long the giant could exist in that specific radii range, to its total lifetime as a giant.
The time-averaged range of  $r_{\rm enc}$ that leads to the binary formation can be found by calculating $r_{\rm form}$ and $r_{\rm wd}$ for each giant's evolutionary moment, and integrating over time. For a BH mass of $7~M_\odot$ and $v_{\infty}=10$km/s (we will use these values as default everywhere thereafter),
  $\overline{r_{\rm form}-r_{\rm wd}}\approx8R_\odot$.
If we increase the mass of the BH, this value slowly increases, but does not exceed  $\sim9.6R_\odot$ for BHs less massive than $100M_\odot$.
This estimate implies that, with a typical RG fraction of about 10\%, typical RG mass of $1 M_\odot$, and core number density of $10^5$ per $pc^3$, one stripped RG can be formed per 25 within-the-core-located BHs  per Gyr.

We find that 80\% of the encounters that form a binary 
take place while the RG is smaller than 10 $R_\odot$, and 50\% of all encounters are taking place when  the RG
is smaller than 4 $R_\odot$ -- the role of the well evolved giants is small,
and subgiants are the most important.
As an example, for a RG of 1$M_\odot$ that has $r_{\rm RG}\sim2~R_\odot$, $r_{\rm form}-r_{\rm wd}\approx5R_\odot$ (where $r_{\rm wd}\sim 3R_\odot$).
However, how much energy is used to directly remove  part of the envelope, 
and how much is simply absorbed by the envelope to make it less bound,
and hence the actual values of $r_{\rm form}-r_{\rm wd}$, 
can only be found by detailed 3D simulations.

\subsection{Three-dimensional hydrodynamical simulations}

\label{sec:3d}

Here we consider encounters of a 7~$M_\odot$BH with a RG that has the mass of 1~$M_\odot$, the radius of $2R_\odot$, and the core of $\sim 0.13~M_\odot$.
To model encounters, we follow the framework described in details in \cite{2014ApJ...786...39N}.
To evolve  the RG  and find the  initial one-dimensional (1D) stellar
profile,  we use the {\tt TWIN/Star}  stellar code  \citep[recent updates
described  in][]{2008A&A...488.1007G}. 
For three-dimensional (3D) simulations, we use the smoothed particle hydrodynamics (SPH) code {\tt STARSMASHER} 
\citep{2010MNRAS.402..105G,2011ApJ...737...49L}.  {\tt STARSMASHER} has been modified in the past
to accept  {\tt TWIN/Star} stellar models \citep{2014ApJ...786...39N}.
We model the 3D star using 50K regular SPH particles and one special particle that represents the core; 
the core mass is $0.13 M_\odot$.
This special particle only interacts gravitationally with other particles.

\begin{figure}
        \includegraphics[width=\columnwidth]{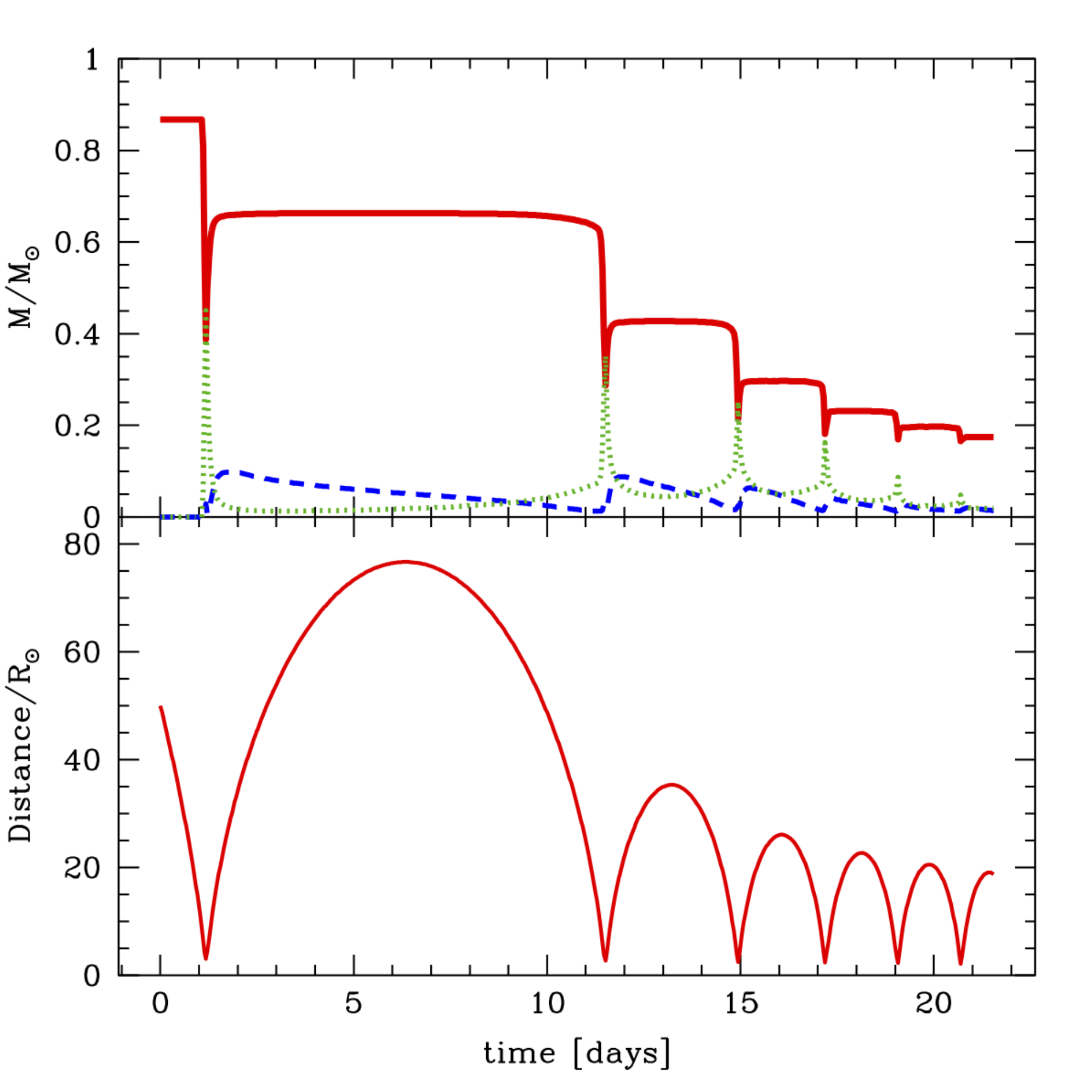}
        \caption{The lower panel shows the instantaneous distance between the RG core and the BH. The upper panel shows the envelope mass that is within the instantaneous RL located around the RG core including other envelope particles bound to that RL (red solid line), the total mass within the RL of the BH excluding the BH itself (blue dashed line), and the mass that is still bound to the binary but is located outside of either of the instantaneous RLs (green dotted line).}
        \label{fig:case3}
\end{figure}

To set up the collision of this star and a $7M_\odot$ BH, 
the two stars are given relative velocity at infinity of $10$ km/s,
and their locations are such that  the resulting periastron distances are
$2.5,3.0,3.25,3.5,3.75,4.0$ and $4.25 R_\odot$.
As we model the entire orbit of the formed binaries, the simulations are very computationally expensive when compared to only modeling the periastron passages. We may expect that increase of the resolution by a factor of two could provide a clarification of the post-encounter binary semi-major axis by a few \% \citep[see, e.g.,][]{2014ApJ...786...39N}. 

A typical simulation is shown in Figure~\ref{fig:case3},
using as an example the case with $r_{p}=3R_\odot$. 
The two stars, after the initial passage, formed a bound binary system.
The semi-major axis in this binary is continuing to decrease.
During the sixth periastron passage, the decrease of the distance between
the RG core and the BH is already only a few percent if compared to the previous passage.
This is because almost all remaining envelope mass is contained within the instantaneous Roche lobe (RL) that the RG has at the periastron. We define the instantaneous RL of the RG (or of the BH) as a 3D equipotential surface that is formed by the RG core itself (or by the BH) and all the particles that are bound to the RG core (or the BH), adopting that the two objects are within an ``instantaneous'' corotating frame that is defined by the separation between the RG center of mass and the BH.  As the cumulative mass of SPH particles bound to the RG core is non-negligible and affects the size of the RL, the exact 3D shape of each RL is found iteratively.
The envelope mass decrease during the sixth passage is $0.019M_\odot$.
Some particles are becoming bound to the BH, but most of them are ejected from the system as the two stars move towards their apastron. By the end of the sixth passage, only about $0.01 M_\odot$ remains bound to the BH.

The formed binary has a semi-major axis of $10.5R_\odot$ and 
an eccentricity of $e=0.8$; the retained envelope mass is $0.174M_\odot$.
The minimum mass that the RG envelope has during the last two  periastron passages is not changing and is $0.17M_\odot$. All that mass is located within $93\%$ of the instantaneous RL that is made by the RG core and the other particles bound to RG core  during the periastron passage.
The distance to the BH during the periastron passage is about 4.7 times larger than the stripped RG. The particles of the envelope therefore are not affect much by tidal perturbations from the BH during the periastron passage.
The binary may be considered formed with the stripped giant having a mass of $0.3M_\odot$.
We note that at the current the radius of the stripped giant is about $0.3R_\odot$, while the same mass was contained within about $0.22R_\odot$ at the start of 3D simulations.
We find that the envelope (that remains bound) has its entropy almost unchanged, as compared to the unperturbed star, for the same mass coordinate.
   
 The model with $r_p=4.25R_\odot$ did not result in a binary formation despite 
the encounter stripping $0.09M_\odot$. 
The model with $r_p=4 R_\odot$ resulted in a formation of a close binary, with  $\sim0.11M_\odot$
ejected during the first periastron passage, and  an initial semi-major axis of about $110 R_\odot$.
The semi-major axis will decrease by a few times, as shown in the model with $r_p=3 R_\odot$ (Fig. 1).
Based on the studies with smaller $r_p$, we estimate that the final minimum amount
of the stripped mass is expected to be at least
$\sim 0.32M_\odot$, as this is the mass that was was outside
of the giant's instantaneous RL yet during the first periastron approach.
As the inner layers expand in all the simulations by $\sim30\%$
as compared to their initial radius coordinate,
the stripped  mass is expected to be $\ga0.4M_\odot$.
It implies that one can expect a creation of stripped
giants with a non-continuous amount of stripped mass:
less than $0.1 M_\odot$ is stripped when a binary is not formed,
and at least $0.4 M_\odot$ is stripped when a binary is formed.

On the other hand, our simulation with the smallest approach at periastron
did not result in a complete envelope removal even after six periastron passages,
where the mass lost during the sixth passage is only $0.008M_\odot$,
and the semi-major axis decreased by only $1.6\%$.
From this it follows that  in our analytic estimate for $r_{\rm wd}$, we either underestimated the energy required for the envelope removal, or overestimated the energy that can be placed into the envelope during a tidal perturbation. If at the periastron the BH is too far away to impose a tidal perturbation on the envelope's layers located close to the core, this part of the envelope is not removed during 3D simulations. The final binary has a semi-major axis of $7.4R_\odot$, an eccentricity of $e=0.77$, and the envelope mass is $0.146M_\odot$. While we do not fine-tune  the exact value of $r_{\rm wd}$ further, it is apparent that the rate of formation of stripped RGs is lower than we estimated in \S~2.1 by a factor of few: from 3D, we have $r_{\rm form}-r_{\rm wd}\ga 1.5R_\odot$, while the simple estimate have predicted  $r_{\rm form}-r_{\rm wd}\approx5R_\odot$; this reduces the total formation rate by a similar factor of 2 to 4.

\section{Properties of the stripped star after an encounter}

To create and evolve stripped remnants, we used the 1D stellar evolution code \texttt{MESA} \citep[Modules for Experiments in Stellar Astrophysics, see instrument papers][]{Paxton2011, Paxton2013, Paxton2015}, revision  8677.
While the structure of the low-mass subgiants obtained with  \texttt{MESA} is very similar to the one obtained with \texttt{TWIN/Star}, only with
{\tt MESA} are we able to strip the subgiant at a mass loss rate comparable to dynamical stripping during the encounter.

\begin{figure}
        \includegraphics[width=\columnwidth]{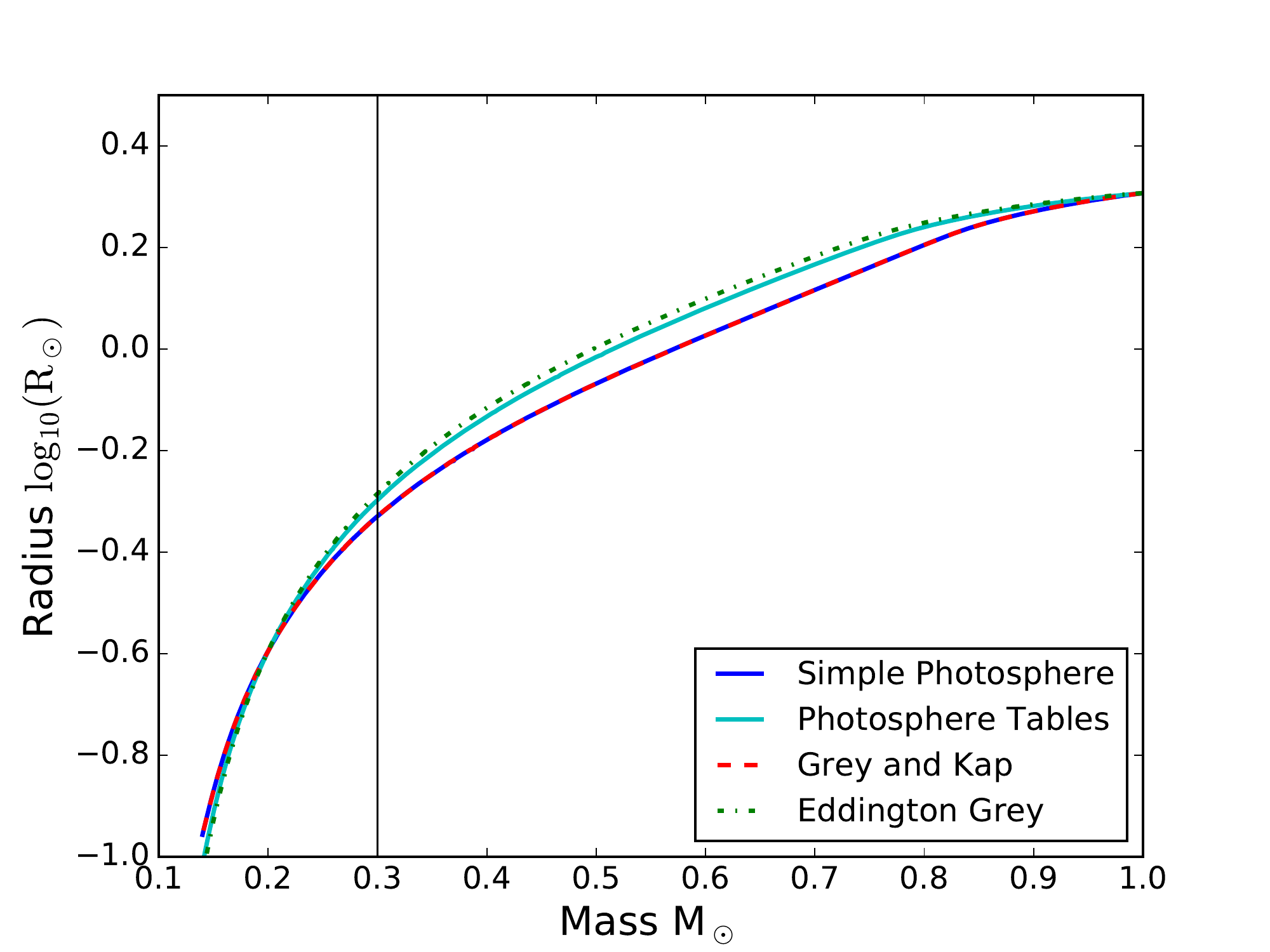}
        \caption{The radii of a $1M_\odot$ and $2R_\odot$ subgiant  while it is evolved with mass loss of $1 M_\odot$ yr$^{-1}$, for four atmospheric boundary conditions.   ``Simple photosphere'' stands for an estimate at the optical depth $\tau=2/3$; in the second choice the precalculated atmosphere tables for photospheres are used; ``Grey and kap'' means that a simple grey atmosphere is calculated to to find consistent pressure $P$, temperature $T$, and opacity at the surface; and ``Eddington Grey'' uses Eddington $T-\tau$ integration} (see {\tt MESA} instrument papers for more details on the used ABCs).
        \label{fig:rad}
\end{figure}

\begin{figure*}
        \includegraphics[width=2.3in]{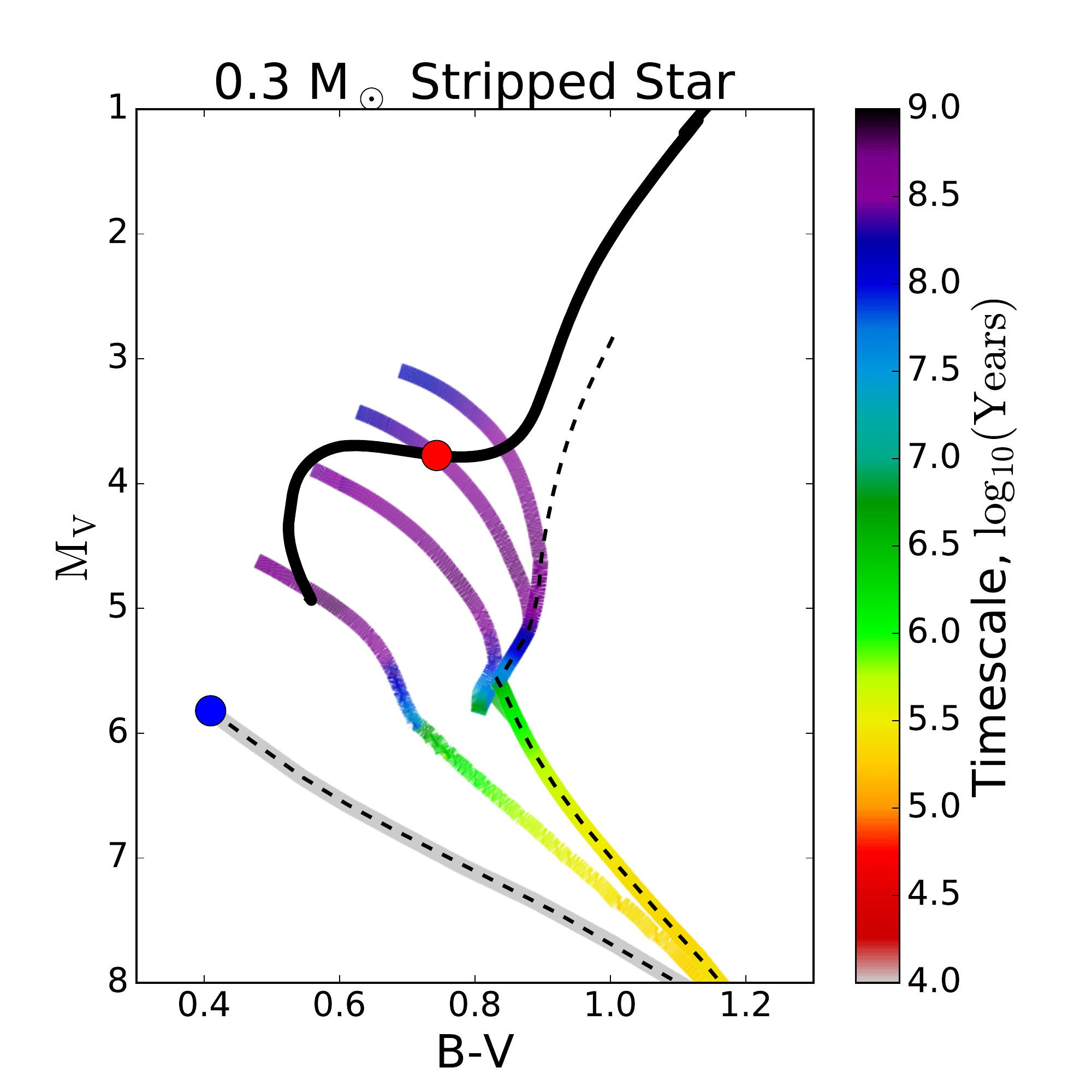}
        \includegraphics[width=2.3in]{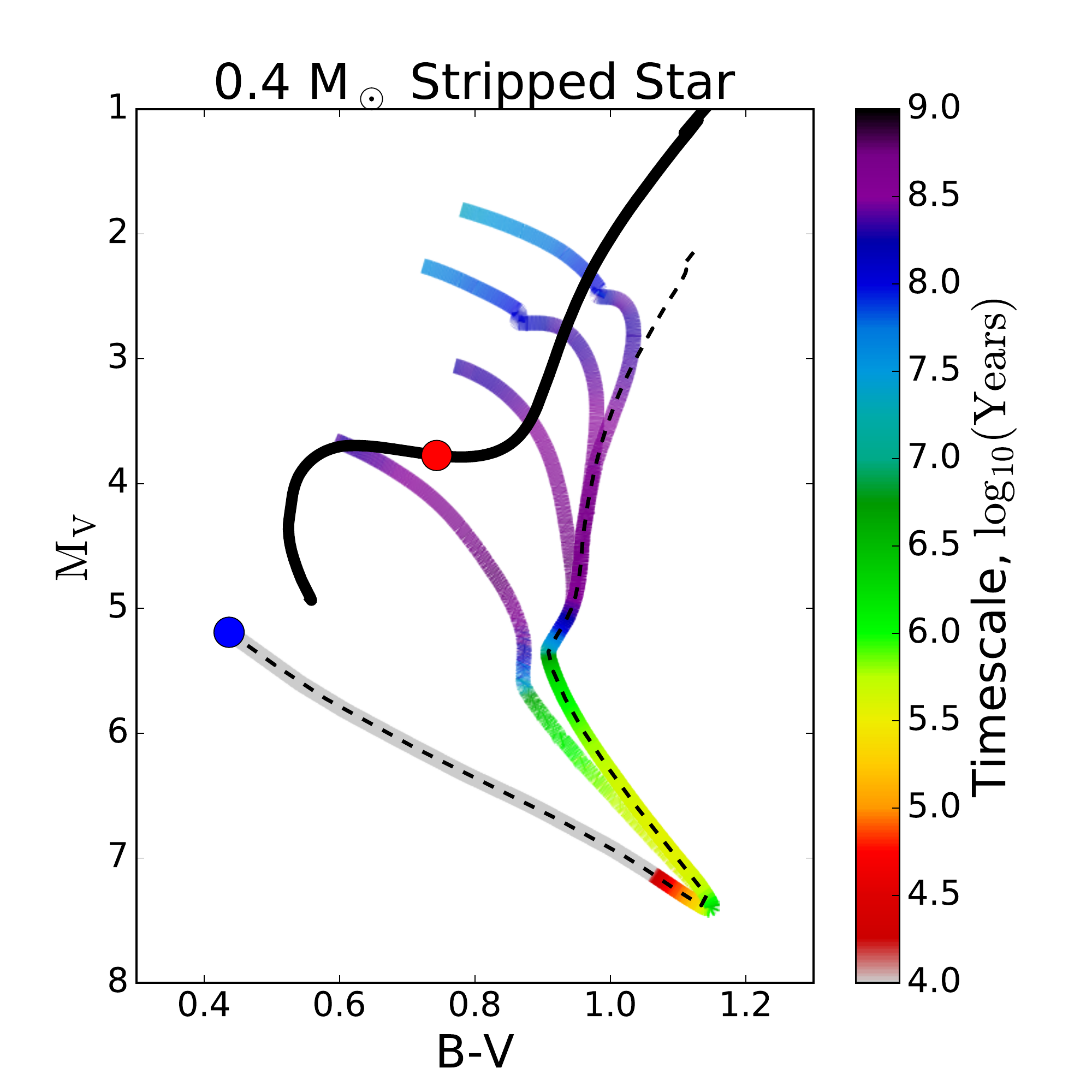}
        \includegraphics[width=2.3in]{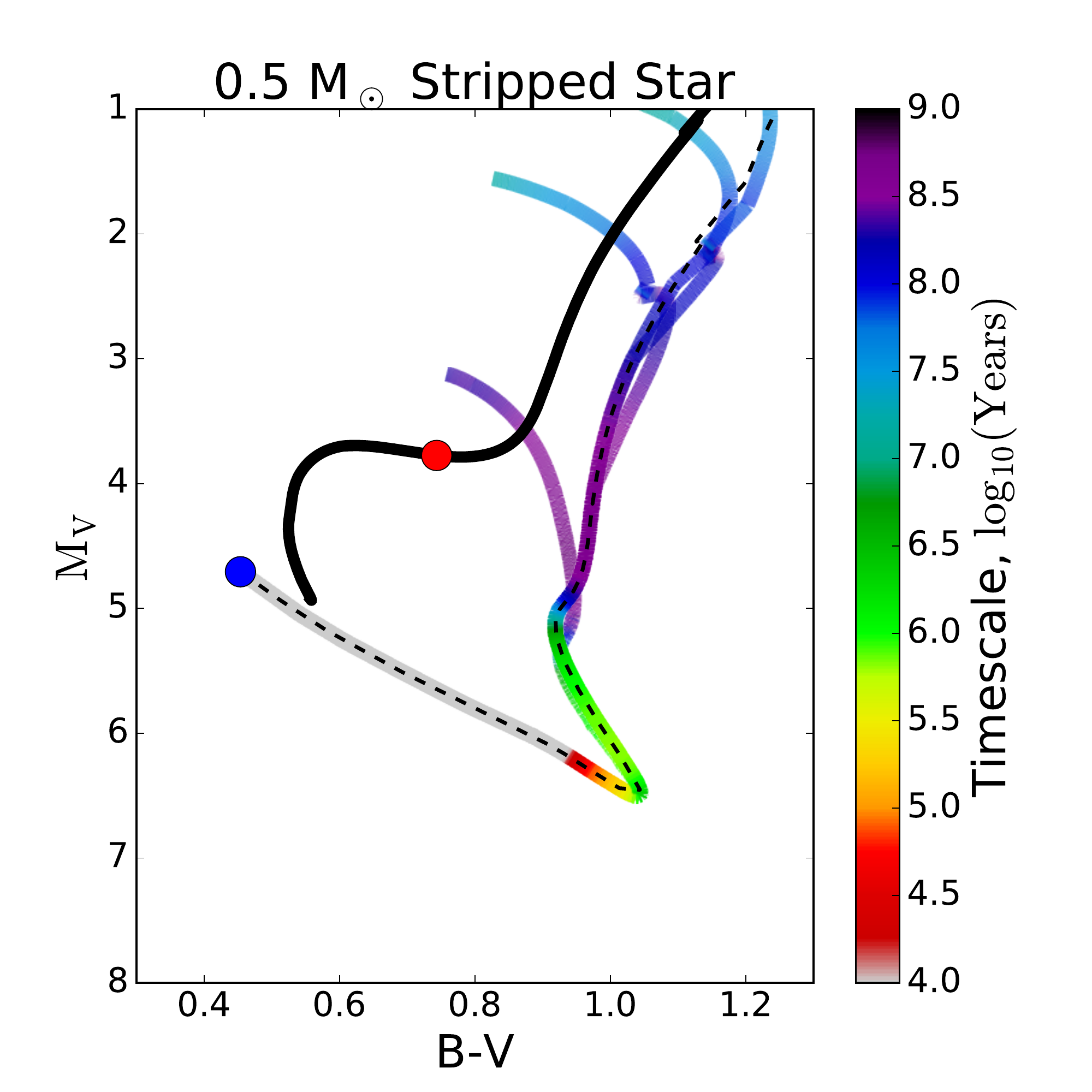}
        \caption{ The evolution of the stripped subgiant. The position of the $1M_\odot$ and $2R_\odot$ subgiant prior the stripping is indicated with the red circle. The location of the stripped star right after the stripping is indicated with the blue circle. The black solid line shows the evolutionary track  of a $1M_\odot$ star.  The black dashed line shows the evolution of the stripped star as it a single star. Tracks with the colors show the evolution during the MT, shown only until the remnant has detached. The colors of the tracks indicate instantaneous characteristic timescales, such that it take to change either the magnitude $M_V$ by one, or color $B-V$ by 0.2.  The left panel shows the case of stripping down to  $0.3M_\odot$
          (the initial orbital periods are $P_{\rm orb}=0.21,0.54,0.97$ and $1.46$ days, from the left to the right), the middle panel shows the case of stripping down to $0.4M_\odot$ ($P_{\rm orb}=0.32,0.88,1.62$ and $2.48$ days), and the right panel shows the case of stripping down to  $0.5M_\odot$ ($P_{\rm orb}= 0.42,1.98,4.22$  and $6.95$ days). }
        \label{fig:cmd03} 
\end{figure*}

We evolved a star with the mass of $1M_\odot$ and metallicity of $Z=0.01$. The evolutionary calculations are done using the default assumptions as in \texttt{MESA}. 
When the star expanded to the radius of $2 R_\odot$, we enforced rapid mass loss of the envelope, at the rate of $1M_\odot$ per year.

 We  found that the radius of the
stripped star during the mass loss, for each given instantaneous total
mass, depends on the atmospheric  boundary conditions (ABCs) that are used, see
Figure~\ref{fig:rad}.
There, we used four ABCs, selected from
  the large  set of conditions  provided with {\tt MESA},  making sure
  that all  of the used  ABCs are  theoretically valid for  studies of
  low-mass  stars,  and  thoroughly  tested  in  previous  studies  of
  low-mass stars by various groups.

Independently from the ABC,
we cannot obtain as small  a radius in our 1D stripped product
as was obtained during an encounter using the 3D code.
The smallest radius that we have obtained with
\texttt{MESA}-stripping for a $0.3M_\odot$
remnant is $0.47R_\odot$, while it is  $0.32R_\odot$ with the 3D code.
We link this different behavior to the entropy of the very outer layers,
with the mass $\le10^{-4}M_\odot$.
The 1D star inside that very outer layer has its entropy unchanged,
like the product of the 3D encounter.
The entropy of this very outer layer in a 1D star has increased compared to its initial value,
by 1 to 45 percent, resulting in star's expansion, while in 3D, the generation of this entropy is not observed. On the one hand, this tiny surface layer might be capable of losing energy radiatively on a timescale of one year, the timescale for the mass loss we used in {\tt MESA}. Hence, it might be expected that the presence of a rarefied but expanded shell of a tiny mass is physically motivated.
  On the other hand, the timescale of the mass loss in 3D is about 10 times faster than in {\tt MESA}, where a purely dynamical mass loss was not possible, and it might be that the observed thermal adjustment would not happen in nature. In any case, dynamically, a layer of this mass cannot affect the shrinkage of the binary.

The binary that is formed in our 3D simulations is highly eccentric.
During the binary periastron approach, the stripped star is at its RL overflow.
We note that the mass of the stripped star is too small
to warrant the transfer of a substantial part of the orbital angular momentum $J_{\rm orb}$ 
into the remaining envelope,
and hence $J_{\rm orb}$ will remain mainly in the binary orbit.
Therefore, if the binary would circularize,
its semi-major axis will be close to $a_{\rm per}(1+e)$, where $e$ is the post-encounter eccentricity.
In principle, it was shown that for the mass ratio as in our system,
both eccentricity and semi-major axis 
may grow in case of impulsive RL overflow at the periastron
\citep{2009ApJ...702.1387S}.
However, we note that a self-consistent treatment
of the MT in eccentric binaries is not yet implemented in {\tt MESA} or other stellar binary codes.

We choose to consider the limiting cases, and to see how different
could be the appearance of the stripped star within those limits.
First, we consider how the stripped star evolves by itself
-- in this case, the remnant will be maximally ``relaxed'' while it evolves after the encounter.
Second, we consider the case of a non-eccentric binary where the MT starts immediately after the mass stripping (the stripped star radius is the same as its RL).
In this case $a=a_{\rm min}=a_{\rm per}$ and the remnant will be evolving as maximally ``perturbed''
(this is not a likely case of the post-encounter evolution due to angular momentum conservation).
Third, we consider the case of a binary that has semi-major axis 
the same as is found from 3D simulations.
This semi-major axis is likely larger than it would be in Nature.
And we consider intermediate cases, with the semi-major axis
chosen between the minimum as above, and the maximum that is found from 3D simulations.
In all the cases, we adopt that the systems are non-eccentric, and hence are evolving to the start of the MT via RL overflow mainly via remnant's expansion and magnetic braking. For magnetic braking, we use default {\tt MESA} prescription for magnetic braking from \cite{Rappaport1983} with $\gamma=4$ .
We adopted that the MT in our binaries is non-conservative, where  the mass accretion rate is limited by the Eddington limit, and the excess mass is lost from the system with the specific angular momentum of the accretor.

The tracks for the $0.3M_\odot$ stripped star are shown in Figure~\ref{fig:cmd03}. Note that the two limiting cases, maximally relaxed and maximally perturbed, restrict the parameter space within which the stripped star can appear on a color magnitude diagram (CMD), and the most likely tracks are in between the two limiting cases. During the MT, the star evolves from the luminosities that are lower than those of the subgiant branch into higher luminosities and hotter temperatures. We define as underluminosity $\delta M_{\rm v}^{\rm und}=M_{\rm V}^{\rm strip}-M_{\rm V}^{\rm norm}$ as the difference between the magnitude of the stripped star, $M_{\rm V}^{\rm strip}$, and an unperturbed star of a similar color, $M_{\rm V}^{\rm norm}$. In our case, normal subgiants have $M_{\rm V}^{\rm norm}\approx3.8$ at $B-V=0.8$. The stripped product has $\delta M_{\rm v}^{\rm und}\ga2$ for a few dozen million years after the stripping, while $1\la M_{\rm v}^{\rm und}\la2$ for  $\sim3\times 10^8$ years.

In encounters with stripping down to $0.4M_\odot$ and $0.5M_\odot$, the formed binaries have slightly larger semi-major axis  than in the case of the star stripped down to $0.3M_\odot$. Similar to the case of $0.3M_\odot$ stripped star,  $0.4M_\odot$ stripped star have $\delta M_{\rm v}^{\rm und}\ga1$ for $\sim1.5\times10^8$ years, but $0.5M_\odot$ stripped star is strongly underluminous  only for less than  $10^7$, if compared to subgiants. However, $0.5M_\odot$ remnants are also redder. They are underluminous by $\delta M_{\rm v}^{\rm und}\ga 2$ for $\sim8\times10^8$ years, if compared to giants that have similar colors (see Figure~\ref{fig:cmd03}).

We verified that
  a similarly long-living strongly underluminous star can be made by stripping of 60\% or more of the donor's mass at a time-averaged $\dot M \ga 10^{-6}M_\odot$yr$^{-1}$. A time-averaged  $\dot M>10^{-6}M_\odot$yr$^{-1}$ from a low-mass subgiant or giant can be supported for the duration of the stripping more than a half of the primary mass only if the donor remains more massive than the companion for most of the mass transfer episode.
A mass transfer with such initial mass ratio is expected to be dynamical unstable.

During the MT, an X-ray binary can appear either as a persistent, or as a transient source. In the first case the X-ray luminosity is expected to be roughly proportional to the mass accretion rate, while a transient source spends most of the time in  quiescence, and has very low X-ray luminosity, typically of the order of $10^{33}$ ergs/s. The bifurcation between appearing as a persistent or a transient source depends on how stable is the accretion disk \citep{2000MNRAS.314..498M,2001A&A...373..251D}.
Using the revised disk instability model \citep{Coriat2012}, 
we find that for most of the time our MT binaries will be in the unstable regime, and hence will
appear as quiescent LMXBs (qLMXBs).

The contribution of the accretion disk to the optical luminosity during quiescence can be estimated using observations of V404 Cyg, the low-mass X-ray binary with an orbital period of 6.47 days and a BH companion of $9\pm^{0.2}_{0.6}M_\odot$  \citep{2016ApJ...818L...5B}. In V404 Cyg, the maximum contribution of the disk in any optical band was estimated to not exceed 14\% of the flux in that band \citep{1996MNRAS.282..977S}. As the most likely value of V404 Cyg's secondary's absolute magnitude is $M_V=3.4$ \citep{1994MNRAS.271L..10S}, the upper limit on the  disk luminosity is thus $M_V\approx 5.5$, and hence the accretion disk potentially can contribute to the total optical luminosity during the first few million years after the encounter.

\section{Conclusions}

We have considered the formation channel of LMXBs in GCs through
grazing encounters between BHs and RGs.
We used a multi-step approach where we first evaluated the total cross-section
  of encounters that can lead to the formation of binaries with a stripped giant companion 
  using the analytical formulae for the tidal energy dissipation and for stellar evolution.
  Second, we clarified the predictions for the range of the closest approaches
  by performing 3D hydrodynamical simulations that used the structure of a $2R_\odot$ giant with a mass of $1M_\odot$ as provided by a fully-fledged 1D stellar evolutionary code.  To find how the product would look like observationally, we evolved the stripped product in a binary using a 1D stellar evolutionary code.

The formation rate is about one binary per Gyr per 50-100 remaining BHs
(if the core number density is $\sim 10^5$ $pc^{-3}$).
Theoretical predictions vary on how many BHs could be present:
from a couple of dozens of BHs in a GC like M22 \citep{2013MNRAS.430L..30S},
to several hundreds of BHs in a cluster that would have a central density about $10^5$ stars per pc$^3$, and a mass of a few times $10^5M_\odot$\citep[e.g., see all models starting with ``n8'' from ][]{2015ApJ...800....9M}.
Our formation channel will be efficiently productive within this predicted range for the retained BHs.

The formed binary unavoidably starts the MT from the captured and stripped subgiant, while
appearing most of the time as a qLMXB.
The donor in such qLMXB is strongly underluminous, by $\delta M_{\rm V}^{\rm und}\ga1$, for a few hundred million years.
  Then after a few hundred million years, our stripped stars will evolve to larger luminosities and will
  blend  with other exotic stars of a GC, e.g. subsubgiants or red stragglers formed via various channels considered in \cite{2017arXiv170310181L}. 
The total timescale of the MT is about $0.5-1$~Gyr.
The expected number of present  qLMXBs is one per $50-200$ BHs retained in a GC core.

We find that the amount of stripped mass in the encounters which did not result in a binary formation, and in those that have resulted in a binary formation, is very different, by at least $0.3M_\odot$.
Neutron stars may also be capable of acquiring companions in a similar manner, while other stars would merge with the subgiant during an encounter that would be close enough for the tidal damping to work, and for the substantial part of the envelope to be ejected.
A time-averaged mass loss rate that can produce as extreme long-living outliers (for the same colors) as our stripped stars is  $\ga10^{-6}M_\odot$yr$^{-1}$, and it has to remove $\ga60\%$ of the donor mass; such mass transfer would not proceed stably in ordinary binaries of that age in the GC. 
Therefore, the presence of a red star that is strongly underluminous, $M_{\rm V}^{\rm und}>1$, may indicate that it has a BH or a NS companion. If a GC qLMXB in a GC  is implied to have a BH accretor, by the ratio of its radio luminosity to its X-ray luminosity, we propose that it is likely to have a donor formed by the considered channel. In about a half of the cases, the donor is expected to remain strongly underluminous, and this is how its counterpart might be identified.

\acknowledgments

NI  acknowledges support  from  CRC program,  and  funding from  NSERC
Discovery.   NI and  KV  acknowledge  that a  part  of  this work  was
performed at  the KITP  which is  supported in part  by the  NSF under
Grant  No.  NSF PHY11-25915.   CAR  acknowledges  for support  Science
without  Borders.  JLAN  acknowledges  CONACyT for  its support.   The
authors thank  the anonymous referees  for the comments  that improved
the clarity of the manuscript.  This  research has been enabled by the
use of computing  resources provided by Compute/Calcul  Canada and the
Shared Hierarchical Academic Research Computing Network (SHARCNET).

\end{document}